\documentclass[secnumarabic,twocolumn,aps,prl,showpacs,showkey,superscriptaddress,prl,floatfix]{revtex4-1}

\usepackage{amsmath,amssymb,amsthm}

\usepackage{hyperref}
\usepackage{amsthm}
\usepackage{graphicx}
\usepackage{subfigure}
\usepackage{color,bbm}
\usepackage[normalem]{ulem}

\definecolor{violet}{rgb}{1.00,0.00,1.00}	

\graphicspath{{./Figures/}}

\newcommand{\R}{\mathbb{R}}

\newcommand{\Exp}[1]{\mathbb{E}\left [ #1 \right]}

\newcommand{\x}{\mathbf{x}}
\newcommand{\J}{\mathbf{J}}
\newcommand{\I}{\mathbf{I}}

\begin{document}

\title{Topological and Dynamical Complexity of Random Neural Networks}

\author{Gilles Wainrib}%
\email[]{wainrib@math.univ-paris13.fr}
\affiliation{LAGA, Universit\'e Paris XIII}
\author{Jonathan Touboul}
\email[]{jonathan.touboul@college-de-france.fr}
\affiliation{The Mathematical Neuroscience Laboratory, CIRB / Coll\`ege de France (CNRS UMR 7241, INSERM U1050, UPMC ED 158, MEMOLIFE PSL*)}
\affiliation{BANG Laboratory, INRIA Paris}

\date{\today}%

\begin{abstract}
	Random neural networks are dynamical descriptions of randomly interconnected neural units. These show a phase transition to chaos as a disorder parameter is increased. The microscopic mechanisms underlying this phase transition are unknown, and similarly to spin-glasses, shall be fundamentally related to the  behavior of the system. In this Letter we investigate the explosion of complexity arising near that phase transition. We show that the mean number of equilibria undergoes a sharp transition from one equilibrium to a very large number scaling exponentially with the dimension on the system. Near criticality, we compute the exponential rate of divergence, called topological complexity. Strikingly, we show that it behaves exactly as the maximal Lyapunov exponent, a classical measure of dynamical complexity. This relationship unravels a microscopic mechanism leading to chaos which we further demonstrate on a simpler® disordered systems, suggesting a deep and underexplored link between topological and dynamical complexity. 
\end{abstract}

\maketitle

Heterogeneity of interconnections is a crucial property to understand the behavior of realistic networks that arise in the modeling of physical, biological or social complex systems. Among these, a paramount example is given by neuronal networks of the brain. In these systems, synaptic connections display characteristic disorder properties~\cite{parker:03,marder-goaillard:06} resulting from development and learning. 
Taking into account this heterogeneity seems now essential, as experimental studies of neuronal tissues have shown that the degree of disorder in the connections significantly impacts the input-output function, rhythmicity and synchrony, effects that can be related to transitions between physiological and pathological behaviors~\cite{aradi-soltesz:02,santhakumar2004plasticity,soltesz2005diversity}. As an example, Aradi and Soltesz~\cite{aradi-soltesz:02} have shown that rats subject to febrile seizure present the same average synaptic properties but increased variance. 

These properties are reminiscent of disordered physical systems such as spin-glasses. The relationship between disorder and qualitative behaviors has been thoroughly studied within theoretical frameworks such as the Sherrington-Kirkpatrick model~\cite{sherrington-kirkpatrick:75} describing the behavior of binary variables interacting through a random connectivity matrix. In these models, estimating the number of metastable states, deemed to be deeply related to the transition to chaos, remains an important endeavor~\cite{parisi2006course}. In the context of nervous system modeling, random neural networks \cite{sompolinsky-crisanti-etal:88, amari:72} constitute a prominent class of models, which describe the evolution of the activity of a neuron $i$ in a $n$-neurons network through the randomly coupled system of ordinary differential equations: 
\begin{equation}
	\label{eq:sompo}
	\dot{x_i} = -x_i + \sum_{j=1}^n J_{ij}S(x_j)
\end{equation}
where $J_{ij}$ are independent centered Gaussian random variables of variance $\sigma^2/n$ representing the synaptic connectivity coefficients between neuron $i$ and $j$~\footnote{{This classical model does not satisfy Dale's principle: a neuron can here be both excitatory and inhibitory.}}, and $S$ is an odd sigmoid function with maximal unit slope at the origin ($S'(x)\leq S'(0)=1$) representing the synaptic nonlinearity. 

The behavior of system \eqref{eq:sompo} has been analyzed in the asymptotic regime of infinite population size $n\to \infty$~\cite{sompolinsky-crisanti-etal:88} and displays a phase transition for a critical value of the disorder at $\sigma=1$: for $\sigma<1$ all the trajectories are attracted to the trivial equilibrium $\x=0$ and for $\sigma>1$ the trajectories have chaotic dynamics. This generic phase transition has been numerically observed in a number of situations in more realistic models involving multiple populations and excitable dynamics~\cite{hermann-touboul:12}. That phase transition, and the chaotic regime beyond the edge of chaos, appear particularly relevant to understand the computational capabilities of neuronal networks. In particular, information processing capacity was characterized as optimal at the edge of criticality~\cite{abbott-toyoizumi:11, sussillo-abbott:09}, and such random neural networks are in particular used in recent machine learning algorithms \cite{jaeger:04}. Moreover, the question of criticality has been widely debated in the theoretical neuroscience community and beyond \cite{bak:96, beggs:08, kitzbichler:09} and it seems that through a number of different mechanisms among which plasticity of synapses, networks may tend to be naturally poised near criticality. In the random neural network, the microscopic mechanisms underpinning this phase transition remains unclear.
It is hence of great interest to dissect precisely the behavior of such systems at the edge of chaos. 

Characterizing the topological modifications of the phase space arising at the edge of chaos is precisely the question we shall address in this Letter. More precisely, we will estimate the averaged number of equilibria in the random neural network~\eqref{eq:sompo}.
To this end, we will develop upon the theories of random matrices~\cite{tao2012topics}, random geometry and Gaussian fields~\cite{adler-taylor:07}. 

We denote by $A_n(\sigma)$ the random number of equilibria (depending on the realization of the matrix $\J=(J_{ij})$). These are the solutions of the system:
\begin{equation}\label{eq:FP}
	x_i = \sum_{j=1}^n J_{ij}S(x_j) 
\end{equation}
For $\sigma<1$, consistently with the mean-field analysis, we first show that $\Exp{A_n(\sigma)} \to 1$ when $n\to \infty$. The proof proceeds by showing that the system is contracting on the whole space $\R^n$. To this purpose, one needs to characterize the eigenvalues of the random Jacobian matrix $-\I+\J.\Delta (S'(\x))$ where $\I$ is the identity matrix, and $\Delta(S'(\x))$ is the diagonal matrix with elements $S'(x_i)$. The matrix $\J.\Delta (S'(\x))$ is a centered Gaussian matrix with independent components, and each column has a distinct standard deviation given by $\sigma^2 S'(x_i)^2/n$. Rajan and Abbott~\cite{rajan-abbott:06} provide the system of equations satisfied by the squared modulus of the eigenvalues of such random matrices, and solve these when considering only two different variances. Their methodology readily generalizes to our problem, and elementary algebraic manipulations (see~\cite{garcia:masterthesis,yiwei}) show that the spectral density has support in the disc of radius $\frac {\sigma^2} n \sum_{i=1}^n{S'(x_i)^2}$ for large $n$. In our case, $S'(x)\leq 1$, and hence all eigenvalues of the matrix $-\I+\J\cdot \Delta(S'(\x))$ have a negative real part in the limit $n\to\infty$.  This implies global contraction of the dynamics, ensuring the fact that the trivial equilibrium $\x=0$ is the unique equilibrium for $\sigma<1$ (Banach fixed point theorem) and its global stability. 

For $\sigma>1$ the situation is more complex. Similarly to the phase transition in spin glasses, the behavior of $\Exp{A_n(\sigma)}$ is likely to scale as $\exp(n\,C(\sigma))$ where $C(\sigma)$ is the {topological complexity}. Our starting point is to observe that the fixed point equations~\eqref{eq:FP} can be viewed as zero crossings of a Gaussian field indexed by $\x\in\R^n$. Moreover, regularity and measurability of that Gaussian field ensure that the Kac-Rice~\footnote{This formula was initially introduced by Kac~\cite{kac:43} to evaluate the average number of real roots of a random polynomial, and a modern account can be found in~\cite{azais:09,adler-taylor:07}} formula can be applied in order to estimate the number of solutions:
\begin{multline*}
	\Exp{A_n(\sigma)} = \int_{\R^n} \textrm{d}\x \;\mathbbm{E}\bigg[ |\det (-\I + \J.\Delta (S'(\x)))| \\
	\times \delta_0(-\x+\J.S(\x))\bigg]. 
\end{multline*}
Recent studies \cite{fyodorov:04,fyodorov-williams:07, auffinger-ben-arous:11} have used that formula to estimate the mean number of critical points in random energy landscapes arising from Hamiltonian systems with strong symmetry properties (spin-glass with spherical symmetry in~\cite{auffinger-ben-arous:11} and translation invariant potential in~\cite{fyodorov:04,fyodorov-williams:07}). In our case, the situation is substantially different: (i) there is no underlying energy landscape since the system \eqref{eq:sompo} is not Hamiltonian and (ii) symmetry properties of the vector field do not enable the same kind of reduction method developed in \cite{auffinger-ben-arous:11,fyodorov-williams:07,fyodorov:04}. A major technical difficulty is the fact that, because the lack of symmetry in our system, one needs to deal with determinant of random matrices with columns of non-identical variances, for which the spectral density is unknown. 

However, near criticality, for $\sigma=1+\varepsilon$ with $0<\varepsilon \ll 1$, we can obtain a first order estimate of the number of equilibria taking advantage of the fact that all equilibria remain close to $\x=0$. More precisely, we first show that with an arbitrarily high probability $1-\xi(n,\varepsilon)$ with $\xi(n,\varepsilon)\to 0$ as $n \to \infty$, all equilibria belong to a ball $\mathcal{B}_{\rho(\varepsilon)}$ centered at $\x=0$ of radius $\rho(\varepsilon)$ which tends to $0$ as $\varepsilon \to 0$. This is a consequence of the spectral analysis of the random matrix $-\I+\J\cdot \Delta(S'(\x))$, whose eigenvalues all have negative real parts outside of $\mathcal{B}_{\rho(\varepsilon)}$ for large $n$. The property that fixed points remain in a small ball around zero is non-trivial. The proof proceeds by defining $\rho(\varepsilon)$ the unique positive solution of the scalar equation $x/\sigma=S(x)$, which is clearly arbitrarily small when $\varepsilon\to 0$, and the smooth modified sigmoid function $S_{\varepsilon,\eta}$ 
\[S_{\varepsilon,\eta}(x)=\begin{cases}
	x/\sigma & \vert x\vert <\rho(\varepsilon)\\
	S(x) & \vert x\vert >\rho(\varepsilon)+\eta
\end{cases}\] 
(the small interval $[\rho(\varepsilon),\rho(\varepsilon)+\eta]$ allows to define a smooth continuation). Because of the properties of the sigmoidal function, and in particular the fact that the differential is decreasing for $x>0$ (and increasing for $x<0$), the same argument as used in the case $\sigma<1$ applied to system~\eqref{eq:sompo} defined with the sigmoid $S_{\varepsilon,\eta}$ (termed \emph{modified system}) ensures that all eigenvalues of $-\I+\J\cdot \Delta(S_{\varepsilon,\eta}'(\x))$ have negative real part and hence that the unique fixed point of the modified system is $0$. In particular, there is no fixed point outside the ball of radius $\rho(\varepsilon)$. In that region the original and modified system are identical, implying that the only possible fixed points of the original system are contained in the ball of radius $\rho(\varepsilon)$.

Therefore, one can split the expectation according to whether $|\x|<\rho(\varepsilon)$ or not, yielding:
\begin{align*}
	\Exp{A_n(\sigma)} =\int_{\mathcal{B}_{\rho(\varepsilon)}} \textrm{d}\x\;\mathbbm{E}\Big[ |\det (-\I + \J.\Delta (S'(\x)))| \\
	\times \delta_0(-\x+\J.S(\x))\Big] + O(\xi(n,\varepsilon))
\end{align*}
Moreover, thanks to the differentiability of the determinant operator, we know that within the ball $\mathcal{B}_{\rho(\varepsilon)}$, the integrand is equal to $|\det (-\I + \J)| + O(\rho(\varepsilon))$, eventually yielding:
\begin{align*}
	\Exp{A_n(\sigma)} =\Exp{|\det (-\I + \J)|} + O(\rho(\varepsilon)+\xi(n,\varepsilon))
\end{align*}
To evaluate this formula, we first compute the logarithm of the determinant:
\begin{align*}
	\frac{1}{n} \log |\det (-\I + \J)| &= \frac 1 n \sum_{\lambda \in sp(\J)} \log |\lambda -1|
	\end{align*} 
where $sp(\J)$ denotes the spectrum of $\J$, which in the large $n$ limit is uniformly distributed in the disc of radius $\sigma$ \cite{girko:85}. Using this property one obtains in the large $n$ limit
\begin{align*}
	\frac{1}{n} \log |\det (-\I + \J)| &= c(\sigma) + R(n)
	\end{align*}
 and  $c(\sigma):=\int_{\mathbb{C}} \log |z -1| \mu_{\sigma}(dz)$ with $\mu_{\sigma}(dz)$ the uniform distribution on the disk of radius $\sigma=1+\varepsilon$ and $R(n)$ is the finite-size error associated to the convergence to the circular law. We have:
\begin{equation}
	\frac 1 n \log \Exp{|\det(-\I+\J)|} = c(\sigma) + \frac 1 n \log \Exp{e^{nR(n)}}.
\end{equation}
This ensures that $\frac 1 n \log \Exp{A_n(\sigma)}$ is arbitrarily close $c(\sigma)$ when $\varepsilon \to 0$ and $n\to\infty$. 

We are hence left computing $c(\sigma)$. Since $\log \vert z\vert$ is harmonic, we can show that for $a,b>0$ :
\[\int_0^{2\pi} \log\vert a - be^{\mathbf{i}\theta}\vert d\theta = 2\pi \log(\min (a, b))\]
ensuring that $c(\sigma)=0$ for $\sigma<1$, which is consistent with our previous analysis of this case, and 
\begin{eqnarray}
	\nonumber c(\sigma)&=&\frac 1 {\pi\sigma^2} \int_1^{\sigma} 2\pi r\log(r)\,dr \\
	&=& \log(\sigma) + \frac 1 2 \left(\frac 1 {\sigma^2}-1\right)\label{eq:CofSigma}
\end{eqnarray}
for $\sigma>1$. For $\sigma$ close to $1^+$, we conclude $c(\sigma)\sim (\sigma-1)^2$. 
Therefore we have shown that:
\begin{equation}\label{eq:result}
	\mathbbm{E}[A_n(\sigma)]\sim e^{n(\sigma-1)^2}
\end{equation}
up to multiplicative polynomial factors.

This combinatorial explosion of the number of equilibria is the hallmark of the accumulation of bifurcations \cite{cessac:95} in a the neighborhood of the critical parameter value $\sigma=1$. Moreover, this result provides a possible topological explanation for the emergence of chaos. For $\sigma>1$, the phase space is heavily mined with equilibrium points, most of which are saddles due to the spectral properties of the Jacobian matrices as discussed above. Typical trajectories evolving in this landscape will hence wander from the vicinity of the different unstable equilibria and attractors appearing, inducing a very high sensitivity to perturbations, distinctive feature of chaos. 

The classical characterization of chaos relies on the evaluation of the maximal Lyapunov exponent of the trajectories quantifying the dynamical complexity. This quantity is defined as follows. Applying
 an infinitesimal external perturbation $x_i\to x_i+\delta x_i^0$ on neuron $i$ at time $t_0$ induces a change on all neurons at subsequent times $x_j(t)\to
 x_j(t)+\delta x_{ji}(t)$ defining a susceptibility matrix $\Psi_{ij}(t_0+\tau,t_0)=\delta x_{ji}(t_0+\tau)/\delta x_i^0$. From the trace of this matrix we shall define an averaged susceptibility $\Psi^2(\tau)=\lim_{t_0\to\infty} \frac 1 n \Exp{\sum_{i,j} \Psi_{ij}^2(t_0+\tau,t_0)}$ and eventually obtain the maximal Lyapunov exponent:
\[\lambda = \lim_{t\to\infty} \frac {\log(\Psi^2(t))} {t}.\]  
This quantity was analyzed in~\cite{sompolinsky-crisanti-etal:88} using a spectral decomposition based on the one-dimensional Schr\"odinger equation. Near criticality, the decomposition dramatically simplifies and yields for $\sigma\sim 1^+$:
\[\lambda \sim {(\sigma-1)^2}.\]
We hence conclude that the topological and dynamical complexities have the same behavior at the edge of chaos.
More surprising is that using the spectral decomposition of the Schr\"odinger equation in the limit $\sigma \gg 1$, Sompolinsky and collaborators show that $\lambda$ diverges as $\log(\sigma)$, precisely as the complexity of the system given in formula~\eqref{eq:CofSigma}, although our analysis rigorously only applies for $\sigma$ close to $1$. 
This further strengthens our microscopic interpretation of the emergence of chaos in relationship with the number of saddles. 

\begin{figure}[htbp]
	\centering
		\subfigure[Lyapunov exponent vs number of fixed point]{\includegraphics[width=.2\textwidth]{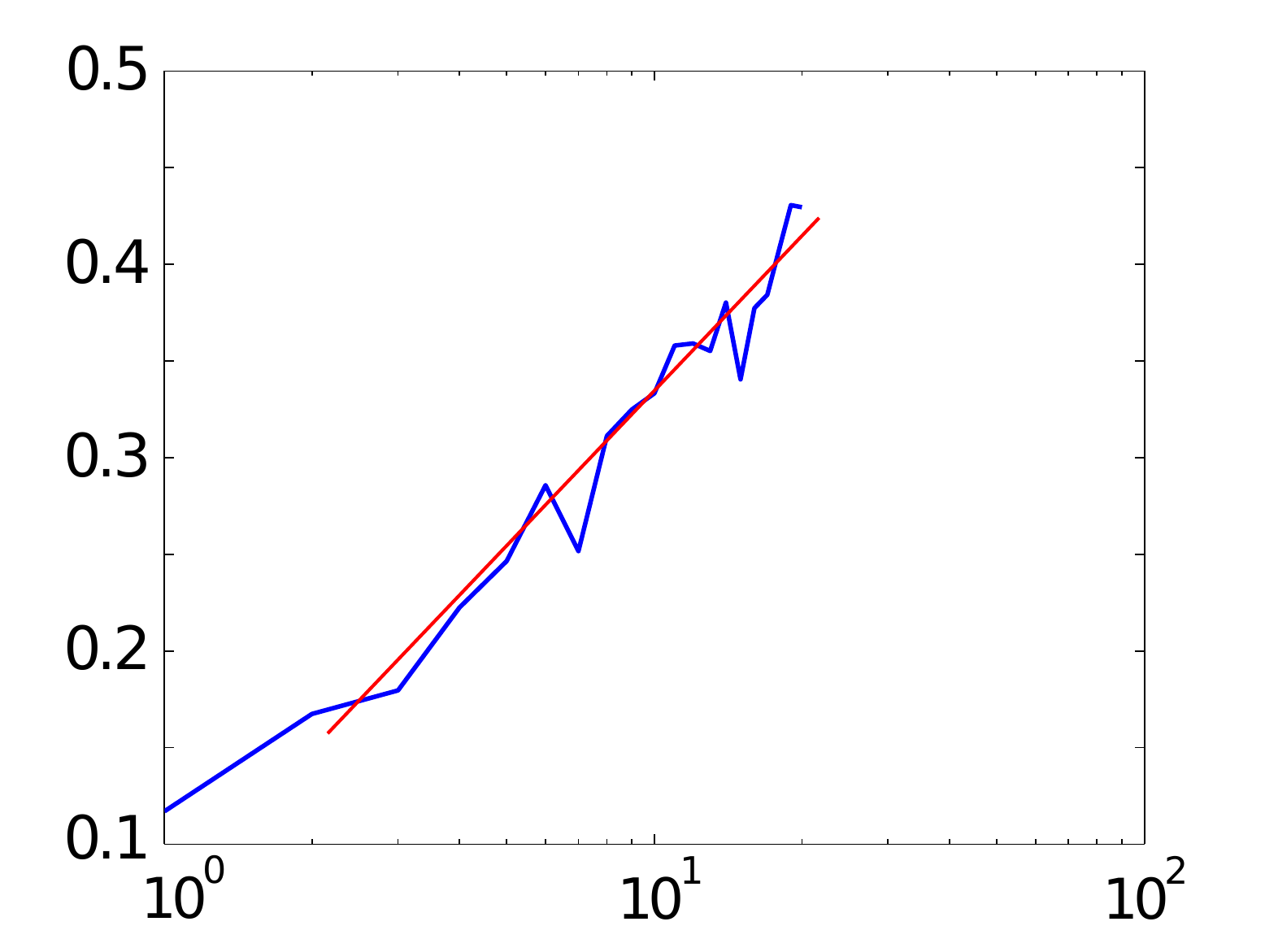}}
		\subfigure[Fakir bed trakectories]{\includegraphics[width=.25\textwidth]{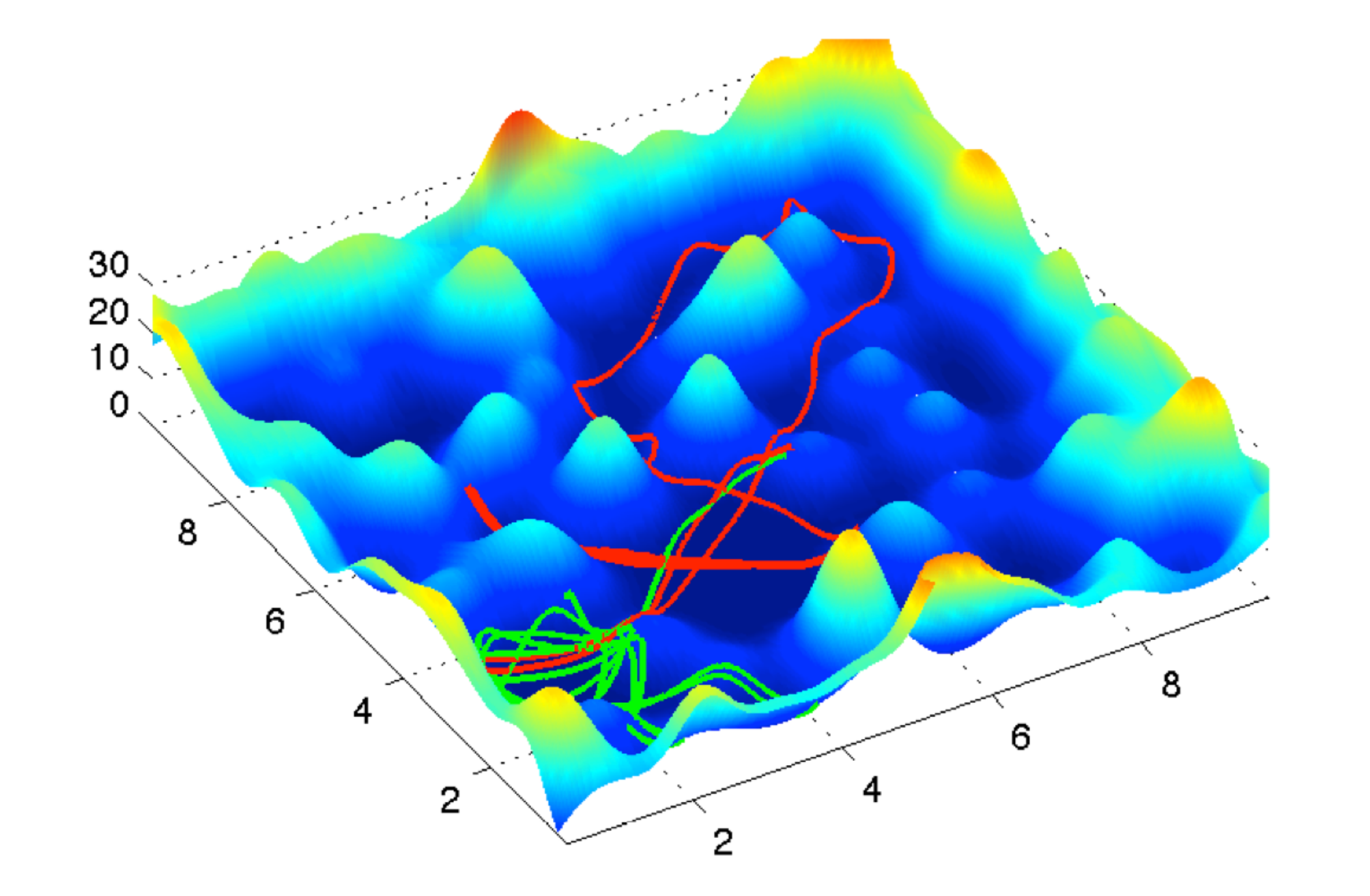}}
	\caption{The Fakir bed dynamics. (a): blue: Lyapunov exponent averaged over 100 realizations of the Fakir's dynamics vs number of fixed points (in semilogarithmic axes) and linear regression (slope $0.36$). (b): Red and Green: two trajectories with very close initial conditions.}
	\label{fig:Fakir}
\end{figure}

If this interpretation is the actual phenomenon arising in random neural networks at the edge of chaos, then the same phenomenon may hold in simpler, lower-dimensional dynamical systems with a large number of unstable fixed points, and these shall display a similar relationship between the number of unstable equilibria and the Lyapunov exponent. Probably the simplest low-dimensional system with easily controllable number of fixed points is the \emph{fakir bed} dynamics, corresponding to the movement of a particle in a two-dimensional complex landscape with $k$ unstable fixed points. More precisely, we now consider a particle confined in a compact subset of $\R^2$, with close to the origin a fixed number $k$ of Gaussian hills (corresponding to the presence of $k$ unstable fixed points) randomly chosen in space. Trajectories of particles soon get chaotic as $k$ increases (see Fig.~\ref{fig:Fakir}(b)), and we numerically compute the maximal Lyapunov exponent of the trajectories. Since the landscape is probabilistic due to the choice of the location of hills, we compute the averaged maximal Lyapunov exponent across $100$ independent realizations of the process, and plot it against the logarithm of the number of equilibria. The corresponding curve (Fig.~\ref{fig:Fakir}(a)), indeed displays an increasing profile well approximated by a linear curve: a clear relationship again emerges between number of unstable fixed points and Lyapunov exponents, supporting our interpretation related to the random neural network.

Thus, motivated by the analysis of fine microscopic phenomena arising at the edge of chaos in random neural networks, we have shown that the average number of equilibria scales exponentially with the system size, with an exponential coefficient proportional to the Lyapunov exponent. This property, relying in part on the spectral theory of random matrices, readily inherits universality properties of the circular law~\cite{tao2008random}, and therefore is valid for a large class of independent couplings beyond Gaussian. Matrices satisfying Dale's principle usually do not fall in this universality class, and do not necessarily present the type of phase transition under consideration. Indeed, Dale's principle typically requires to consider correlated or non-centered synaptic weights, generally modifying the spectrum of the connectivity matrix and the network dynamics. Combining recent results on the spectrum of such matrices~\cite{rajan-abbott:06,tao2011outliers} with the Kac-Rice formula should provide new insight into the complexity of these networks. 

Moreover, our result shows that the complexity smoothly with the disorder parameter with a critical exponent $2$, larger than $1$, which corresponds to second order phase transitions. This property ensures a form of structural robustness in the neighborhood of the phase transition in the sense that the complexity-related properties of the critical state would hold beyond the edge of chaos, which may have several implications in information processing capabilities~\cite{abbott-toyoizumi:11}. Moreover, the result obtained~\eqref{eq:result} reveals a particular scaling $\sigma=1+O(n^{-\frac 1 2 })$ characterizing the typical thickness of the edge of chaos for large finite-size networks. This study is the first application to random neural networks of recent methods used for counting the number of metastable equilibria in spin glasses. From the theoretical viewpoint, we extended that approach to out of equilibrium, non-Hamiltonian systems at zero temperature (singular points of a vector field). This identity incidentally found between topological and dynamical complexity highlights what we conjecture to be a deep correspondence in large complex systems. Numerous questions and perspectives have emerged from this study, among which the estimation of the distribution of equilibria and their number beyond the edge of criticality, requiring significant advances in the analysis of random matrix determinants, or pursuing the exploration, in line with \cite{ledrappier1985metric,frederickson1983liapunov,chatterjee:08}, of the relationship between topological and dynamical complexity with other measures such as the fractal dimensions of chaotic attractors.
\medskip

\end{document}